\documentstyle[aps,twocolumn,prb,aps]{revtex}

\begin{document}

\wideabs{
\title{{\bf Nonlinear parametric instability in double-well lattices}}
\author{Jaroslav Riedel$^{1}$, Boris A. Malomed$^{2}$, and Eva Majern\'{\i}kov\'{a}$
^{1}$}
\address{$^1$Department of Theoretical Physics, Palack\'{y} University,\\
T\v{r}. 17. Listopadu 50, 77207 Olomouc, Czech Republic\\
$^2$Department of Interdisciplinary Studies, Faculty of Engineering,\\
Tel Aviv University, Tel Aviv 69978, Israel}
\maketitle

\begin{abstract}
A possibility of a nonlinear resonant instability of uniform oscillations in
dynamical lattices with harmonic intersite coupling and onsite nonlinearity
is predicted. Numerical simulations of a lattice with
a double-well onsite anharmonic potential confirm the existence of the
nonlinear instability with an anomalous value of the corresponding power
index, $\approx 1.57$, which is intermediate between the values
$1$ and $2$ characterizing the linear and nonlinear (quadratic)
instabilities. The anomalous power index may be a result
of a competition
between the resonant quadratic instability and nonresonant linear
instabilities. The observed instability
triggers transition of the lattice into a chaotic dynamical state.
\end{abstract}
}

\narrowtext
Dynamical lattices with onsite nonlinearity and harmonic intersite coupling
constitute a vast class of models which have numerous physical application
and are an object of great interest in their own right, see, e.g., Ref. \cite
{general}. Among these models, the ones with a {\it double-well} onsite
potential [given, e.g., by the expression (\ref{potential}) below] have
special importance, as they directly apply to the description of structural
transitions in dielectrics, semiconductors, superconductors, and optical
lattices (see recent works \cite{doublewell} and references therein), and
find other applications \cite{others}.

The simplest dynamical state in lattices represents spatially homogeneous
oscillations. This state in conservative lattice models is sometimes stable,
and sometimes it is subject to linear modulational instabilities initiating
a transition to nontrivial dynamics \cite{MI}. An objective of the present
work is to demonstrate analytically, and verify by direct simulations, that
homogeneous oscillatory states in lattices with the double-well
anharmonicity may be subject to a specific {\em nonlinear} instability,
which triggers transition of the lattice into a chaotic dynamical state.

The nonlinear parametric instability, which is a subject of this work, is
inherently related to the phonon anharmonism in the lattice: the instability
is caused by a resonance involving the uniform oscillations of the lattice
and a second harmonic of the phonon mode. A possibility of a nonlinear
instability of an {\it intrinsic localized mode} \cite{IM}\ in the lattice
due to the phonon anharmonism was first considered in Ref. \cite{me}. An
opposite, and more common, type of the resonance, namely, between a strictly
linear phonon mode and a higher harmonic of an intrinsic localized mode is
well known to give rise to a slow decay of the localized mode into phonons
(see recent works \cite{Aubry} and references therein).

The general form of the lattice equation of motion is 
\begin{equation}
\ddot{u}_{n}+f\left( u_{n}\right) u_{n}=u_{n+1}+u_{n-1}-2u_{n},
\label{general}
\end{equation}
where $u_{n}$ are real dynamical variables on the lattice, the overdot
stands for $d/dt$, $f\left( u_{n}\right) $ is a polynomial function
accounting for the onsite nonlinearity (in fact, nonpolynomial functions can
be considered too), and the right-hand side of the equation accounts for the
intersite harmonic coupling. The linearized version of Eq. (\ref{general})
gives rise to phonon modes 
\begin{equation}
u_{n}=A\sin \left( kn-\omega t\right)  \label{mode}
\end{equation}
with an arbitrary infinitesimal amplitude $A$ and the dispersion relation $%
\omega =2\left| \sin (k/2)\right| $ inside the phonon band, $\omega \leq 2$
(by definition, the frequencies are positive).

A homogeneous oscillatory state $U_{0}(t)$ is a time-periodic solution to
the equation 
\begin{equation}
\ddot{U}_{0}+f\left( U_{0}\right) U_{0}=0  \label{U0}
\end{equation}
with a fundamental frequency $\Omega $. The linear stability of the
homogeneous state is determined by a linearized equation for small
perturbations $\delta u_{n}$, which is produced by the substitution of $%
u_{n}=U_{0}(t)+\delta u_{n}$ into Eq. (\ref{general}): 
\begin{eqnarray}
&&\delta \ddot{u}_{n}+\left[ f^{\,\,\prime }\left( U_{0}(t)\right)
U_{0}(t)+f\left( U_{0}(t)\right) \right] \delta u_{n}  \nonumber \\
&=&\delta u_{n+1}+\delta u_{n-1}-2\delta u_{n}\,.  \label{linearized}
\end{eqnarray}
\newline
In the mean-field approximation, one may describe phonon modes of the type (%
\ref{mode}) on top of the homogeneous oscillations, replacing the
coefficient in front of $\delta u_{n}$ on the left-hand side of Eq. (\ref
{linearized}) by its time-average value $\omega _{0}^{2}\equiv \left\langle
f^{\prime }\left( U_{0}(t)\right) U_{0}(t)+f\left( U_{0}(t)\right)
\right\rangle $. The corresponding dispersion relation for the phonon modes
acquires a {\it gap} $\omega _{0}$, so that 
\begin{equation}
\omega ^{2}=\omega _{0}^{2}+4\sin ^{2}(k/2),  \label{dispersion}
\end{equation}
which gives rise to the phonon band 
\begin{equation}
\omega _{0}^{2}\leq \omega ^{2}\leq 4+\omega _{0}^{2}  \label{band}
\end{equation}
(if $\omega _{0}^{2}<0$, the homogeneous oscillations are immediately
unstable).

Beyond this simple approximation, the Fourier decomposition of the
coefficient $f^{\,\,\prime }\left( U_{0}(t)\right) U_{0}(t)+f\left(
U_{0}(t)\right) $ in Eq. (\ref{linearized}) gives rise to parametrically
driven terms $\,\sim \cos (m\Omega t)\cdot \delta u_{n}$ with all integer
values of $m$. The linear parametric drive resonates with a perturbation
frequency $\omega $, i.e., it may give rise to a {\em resonant} linear
instability, under the condition $m\Omega -\omega =\omega $, or 
\begin{equation}
\omega =\omega _{{\rm res}}^{{\rm (lin)}}\equiv (m/2)\Omega .
\label{linearRes}
\end{equation}
If any resonant frequency $(m/2)\Omega $ gets into the phonon band (\ref
{band}), the homogeneous oscillatory state is expected to be modulationally
unstable, otherwise the resonant linear instability does not take place.

In the latter case, it makes sense to seek for nonlinear parametric
instabilities, the simplest of which may be generated by a cubic term, or
any higher-order one, in the onsite nonlinearity. Indeed, the cubic term
generates a nonlinear correction $\sim U_{0}(t)\cdot \left( \delta
u_{n}\right) ^{2}$ to Eq. (\ref{linearized}), which may be regarded as a
parametric drive that can give rise to a {\em nonlinear} parametric
resonance under the condition $m\Omega -2\omega =\omega $, or 
\begin{equation}
\omega =\omega _{{\rm res}}^{{\rm (nonlin)}}\equiv (m/3)\Omega ,
\label{nonlinRes}
\end{equation}
where $m$ is an arbitrary integer different from a multiple of $3$, cf. Eq. (%
\ref{linearRes}) (if $m$ is a multiple of $3$, the linear parametric
resonance takes place at the same frequency, so that the nonlinear resonance
is insignificant).

Of course, this description has a very approximate nature for two reasons.
First, the phonon band (\ref{band}) was defined in the framework of the
mean-field approximation, hence one cannot be sure in the accuracy of the
predictions based on the comparison of the resonant frequencies with this
band. Second, the full lattice model (\ref{general}) may give rise to other
instabilities, which are not related to the parametric resonance. Therefore,
the above consideration should only be considered as a qualitative clue, and
an actual possibility of dynamical regimes dominated by the nonlinear
resonance must be checked by direct simulations.

Continuing the consideration, we note that, if none of the linear-resonance
frequencies (\ref{linearRes}) gets into the renormalized band (\ref{band}),
but a nonlinear-resonance frequency (\ref{nonlinRes}) can be found inside
the band, an evolution equation for the amplitude of the corresponding
resonant-perturbation mode, $\delta u_{n}=A(t)\cos \left( \omega _{{\rm res}%
}t\right) \cdot v_{n}$ , with some spatial profile $v_{n}$ (it may be, for
instance, the above-mentioned localized intrinsic mode), has a general form 
\begin{equation}
dA/dt=CA^{2},  \label{nonlinear}
\end{equation}
where $C$ is a constant which depends on a particular form of Eq. (\ref
{general}) and the homogeneous solution $U_{0}(t)$; cf. a similar equation
governing the nonlinear instability of the so-called {\it embedded solitons} 
\cite{embedded}. A solution to Eq. (\ref{nonlinear}) is 
\begin{equation}
A=A_{0}/\left( 1-CA_{0}t\right) ,  \label{growth}
\end{equation}
where $A_{0}$ is the initial value of the perturbation amplitude. A drastic
difference of the perturbation growth law (\ref{growth}) from the
exponential growth in the case of the linear instability is that the
nonlinear instability is initially growing much slower than an exponential,
and a characteristic time scale of the growth, $\sim 1/\left( CA_{0}\right) $%
, depends on the initial perturbation $A_{0}$, while in the case of the
exponential growth it is a fixed constant. However, the nonlinear
instability is self-accelerating, and, as a manifestation of that, Eq. (\ref
{growth}) formally predicts a singularity at $t=1/\left( CA_{0}\right) $. In
reality, of course, the singularity may not occur, as the above
approximation, taking into regard the first nonlinear correction to Eq. (\ref
{linearized}), becomes irrelevant if $A(t)$ is too large. A natural
conjecture, that will be corroborated by direct simulations below, is that
the nonlinear instability leads to a chaotic dynamical state.

It is relevant to mention that, although nonlinear instabilities are less
common than the usual linear instability, they occur and play an important
role in many physical problems, as diverse as optical solitons in media with
competing quadratic and cubic nonlinearities \cite{embedded}, Bose gases,
plasma turbulence, contact lines in flows, etc. \cite{nonlininstab}. In this
work, we will check the possibility of the nonlinear instability of the
homogeneous oscillations in the lattice model (\ref{general}) with 
\begin{equation}
f(u_{n})=-u_{n}^{2}+\nu u_{n}^{4},\,\nu >0,  \label{f}
\end{equation}
which corresponds to the double-well onsite anharmonic potential, 
\begin{equation}
V(u_{n})=-u_{n}^{4}/4+\nu u_{n}^{6}/6.  \label{potential}
\end{equation}
In this case, Eq. (\ref{U0}) can be solved in terms of elliptic functions,
but an explicit result is very cumbersome.

As a typical example, we take homogeneous oscillations produced by Eq. (\ref
{U0}) with $\nu =0.01$ and initial conditions $U_{0}(0)=1$ and $\stackrel{.}{%
U}_{0}(0)=0$. The variable $U_{0}$ then performs strongly anharmonic
oscillations between the values $\left( U_{0}\right) _{\min }=1$ and $\left(
U_{0}\right) _{\max }=12.247$ at the fundamental frequency $\Omega
=\allowbreak 1.\,\allowbreak 694$, and the gap in the renormalized phonon
spectrum (\ref{dispersion}) is calculated to be $\omega _{0}=\allowbreak
3.\,518$, so that the renormalized band (\ref{band}) is, in the mean-field
approximation, 
\begin{equation}
3.518<\omega <4.\allowbreak 047.  \label{realband}
\end{equation}
Then, it is straightforward to check that all the linear-resonance
frequencies (\ref{linearRes}) do {\em not} get into this band (the band as
whole fits between the linear resonant frequencies $3.392$ and $4.240$,
which correspond to $m=4$ and $m=5$). On the other hand, the
nonlinear-resonance frequency (\ref{nonlinRes}) corresponding to $m=7$ is $%
\omega _{{\rm res}}^{{\rm (nonlin)}}=\allowbreak 3.\,\allowbreak 957\,3$,
which lies {\em inside} the band (\ref{realband}) [all the other frequencies
given by Eq. (\ref{realband}) are located outside the band].

To directly test the instability, small perturbations of the form 
\begin{equation}
\delta u_{n}(0)=A_{0}\cos (2\pi p_{0}n/N),  \label{p}
\end{equation}
where $N$ is the net number of sites in the lattice and $p_{0}$ is an
integer, were added to the homogeneous oscillatory state. The lattice
equations of motion were solved for $N=1000$ and periodic boundary
conditions by means of the eighth-order explicit Runge-Kutta scheme with a
stepsize control such that the time step was dynamically changed within the
range $0.05$ - $0.3$. It was checked that the relative (per site) error at
each step did not exceed $10^{-10}$.

The simulations were performed for the perturbations (\ref{p}) with $p_{0}$
taking values in the interval $1\leq p_{0}\leq 30$. In all the cases
considered, results were quite similar. Here, we demonstrate a typical
example with $p_{0}=20$. Long-time evolution initiated by the small
perturbation (\ref{p}) with $A_{0}=0.05$ is displayed is Fig. 1 in the form
of a set of plots showing the temporal development of several components in
the Fourier transform of the lattice field, which are defined as follows: 
\begin{eqnarray}
U_{p}(t) &=&\left( 2/N\right) \sum_{n=1}^{N}u_{n}(t)\exp \left( 2i\pi
pn/N\right) ,\,p\neq 0;  \nonumber \\
U_{0}(t) &=&\left( 1/N\right) \sum_{n=1}^{N}u_{n}(t)\,.  \label{Fourier}
\end{eqnarray}
It is obvious that the small perturbation triggers a transition of the
lattice into a chaotic state. Fully developed chaos, i.e., a state in which
all the lattice modes are involved into the chaotic motion, is attained at $%
t\approx 22$, when the phonon mode with $p=p_{0}+1$ gets chaotically excited
too, see Fig. 1. To further illustrate the transition to chaos, in Fig. 2 we
additionally show in detail, on the logarithmic scale, the growth of the
amplitude $|U_{p_{0}+1}(t)|$. More detailed studies of the established
chaotic state may be of interest in their own right, but this problem is
beyond the scope of the present work.

As concerns the nonlinear character of the instability, a crucial issue is
the growth of the perturbation at the initial stage. It is necessary to
check whether it is indeed essentially different from the familiar
exponential law, being, instead, close to the Eq. (\ref{growth}). To this
end, in Fig. 3 we display the best fit of the time evolution of the
numerically computed Fourier amplitude $|U_{p_{0}}(t)|$ to a function 
\begin{equation}
A_{{\rm fit}}(t)=\Delta \cdot (1-\gamma t)^{-\alpha },  \label{fit}
\end{equation}
where the parameters are found to be $\Delta =0.041$, $\gamma =0.560$ and $%
\alpha =1.750$.

Comparison of these results with Eq. (\ref{growth}) shows a difference in
the (most essential) power parameter $\alpha $. Note that the expression ( 
\ref{fit}) with the empirically found value $\alpha =1.750$ formally
corresponds to a solution to the nonlinear evolution equation $%
dA/dt=CA\allowbreak ^{\beta }$, with an anomalous value of the power index, $%
\beta \equiv 1+\alpha ^{-1}\approx 1.5714$, that should be compared to Eq. ( 
\ref{nonlinear}), valid in case of the ordinary nonlinear instability \cite
{embedded}. This anomalous value is sort of intermediate between $\beta =1$
and $\beta =2$, which are expected for the linear and nonlinear
instabilities, respectively. This result may suggest that, in fact, in the
present model we have a competition between the resonant nonlinear
instability, qualitatively considered above, and linear instabilities
against {\em nonresonant} perturbations, which were not taken into regard in
the above consideration. While an accurate analysis of the full linear
stability problem of the homogeneous oscillations is a technically complex
problem, that we do not aim to consider here, Fig. 3 clearly shows that the
resonant nonlinear instability dominates in the growth of the perturbations.

In conclusion, we have proposed a possibility of a nonlinear resonant
instability of homogeneous oscillations in harmonically coupled nonlinear
lattices, which is expected to play a dominant role, provided that no
resonant frequency accounting for the linear parametric resonant instability
gets into the renormalized phonon band, while a frequency that gives rise to
a quadratic parametric resonance is found in the band. Numerical simulations
of the lattice with a double-well onsite anharmonic potential confirm the
existence of nonlinear instability with an anomalous value of the power
index $\approx 1.57$, which is intermediate between the values $1$ and $2$,
characteristic of the linear and nonlinear instabilities. The onset of the
nonlinear instability triggers transition of the lattice into a chaotic
dynamical state.

A valuable discussion with P.G. Kevrekidis is acknowledged. B.A.M.
appreciates hospitality of the Department of Theoretical Physics at the
Palack\'{y} University (Olomouc, the Czech Republic). E.M. acknowledges a
partial support by grants No. 202/01/1450 from the agency GACR and by VEGA
No. 2/7174/20.

\newpage

\section*{Figure Captions}

Fig. 1. The time dependence for selected Fourier amplitudes $|U_{p}(t)|$,
defined as per Eq. (\ref{Fourier}). The results are shown for $p=0$, $%
p=p_{0} $, $p=2p_{0}$, and $p=p_{0}+1$, where $p_{0}=20$. The lattice size
is $N=1000 $ (with periodic boundary conditions), and the initial amplitude
of the perturbation is $A_{0}=0.05$.

Fig. 2. Details of the evolution of the Fourier amplitude $|U_{p_{0}+1}(t)|$%
, shown on the logarithmic scale.

Fig. 3. Fitting the time dependence of the amplitude $|U_{p_{0}}(t)|$ to the
function (\ref{fit}) with $\Delta =0.0411$, $\gamma =0.560$ and $\alpha
=1.750$. Diamonds stand for numerical data, and stars (which almost
completely overlap with the diamonds) show the closest values provided by
the fitting function.

\end{document}